\documentclass[a4paper,10pt]{article}
\usepackage{graphicx}
\usepackage{axodraw4j}
\usepackage{amssymb}
\usepackage{amsmath}
\usepackage{cite}
\usepackage{cite}
\usepackage{graphicx,subfigure,epsfig,epsf}
\def\be{\begin{equation}}
\def\ee{\end{equation}}
\def\bea{\begin{eqnarray}}
\def\eea{\end{eqnarray}}

\def\issue(#1,#2,#3){#1 (#3) #2} 
\def\APP(#1,#2,#3){Acta Phys.\ Polon.\ \issue(#1,#2,#3)}
\def\ARNPS(#1,#2,#3){Ann.\ Rev.\ Nucl.\ Part.\ Sci.\ \issue(#1,#2,#3)}
\def\CPC(#1,#2,#3){Comp.\ Phys.\ Comm.\ \issue(#1,#2,#3)}
\def\CIP(#1,#2,#3){Comput.\ Phys.\ \issue(#1,#2,#3)}
\def\EPJC(#1,#2,#3){Eur.\ Phys.\ J.\ C\ \issue(#1,#2,#3)}
\def\EPJD(#1,#2,#3){Eur.\ Phys.\ J. Direct\ C\ \issue(#1,#2,#3)}
\def\IEEETNS(#1,#2,#3){IEEE Trans.\ Nucl.\ Sci.\ \issue(#1,#2,#3)}
\def\IJMP(#1,#2,#3){Int.\ J.\ Mod.\ Phys. \issue(#1,#2,#3)}
\def\JHEP(#1,#2,#3){J.\ High Energy Physics \issue(#1,#2,#3)}
\def\JPG(#1,#2,#3){J.\ Phys.\ G \issue(#1,#2,#3)}
\def\MPL(#1,#2,#3){Mod.\ Phys.\ Lett.\ \issue(#1,#2,#3)}
\def\NP(#1,#2,#3){Nucl.\ Phys.\ \issue(#1,#2,#3)}
\def\NIM(#1,#2,#3){Nucl.\ Instrum.\ Meth.\ \issue(#1,#2,#3)}
\def\PL(#1,#2,#3){Phys.\ Lett.\ \issue(#1,#2,#3)}
\def\PRD(#1,#2,#3){Phys.\ Rev.\ D \issue(#1,#2,#3)}
\def\PRL(#1,#2,#3){Phys.\ Rev.\ Lett.\ \issue(#1,#2,#3)}
\def\PTP(#1,#2,#3){Progs.\ Theo.\ Phys. \ \issue(#1,#2,#3)}
\def\RMP(#1,#2,#3){Rev.\ Mod.\ Phys.\ \issue(#1,#2,#3)}
\def\SJNP(#1,#2,#3){Sov.\ J. Nucl.\ Phys.\ \issue(#1,#2,#3)}

\def\d{\delta}
\def\ep{\varepsilon}
\def\no{\nonumber}
\def\uno{\mbox{1 \kern-.59em {\rm l}}}
 
\begin{document}
\bibliographystyle{unsrt}
\begin{titlepage}

\vskip2.5cm
\begin{center}
\vspace*{5mm} {\huge \LARGE Cosmic microwave background
polarization in Noncommutative space-time}
\end{center}
\vskip0.2cm

\begin{center}
{\it S. Batebi \footnotemark[1], M. Haghighat \footnotemark[2],
R. Mohammadi \footnotemark[3], S. Tizchang \footnotemark[1]}

\end{center}
\vskip 8pt

\begin{center}

$\footnotemark[1]$ {\it  Department of Physics, Isfahan University of Technology,\\ Isfahan 84156-83111, Iran}\\
\vspace*{0.3cm}
$\footnotemark[2]$ {\it  Department of Physics, Shiraz University ,\\Shiraz, Iran}\\
$\footnotemark[3]$ {\it Iran Science and Technology Museum (IRSTM), PO BOX: 11369-14611, Tehran, Iran.}\\
\vspace*{0.3cm}


\end{center}


\begin{abstract}
In the standard model of cosmology (SMC) the B-mode polarization
of the CMB can be explained by the gravitational effects in the
inflation epoch.  However, this is not the only way to explain the
B-mode polarization for the CMB.  It can be shown that the Compton
scattering in presence of a background besides generating a
circularly polarized microwave, can leads to a B-mode polarization
for the CMB.  Here we consider the non-commutative (NC) space time
as a background to explore the CMB polarization at the last
scattering surface.  We obtain the B-mode spectrum of the CMB
radiation by scalar perturbation of metric via a correction on the
Compton scattering in NC-space-time in terms of  the circular
polarization power spectrum and the non-commutative energy scale.
It can be shown that even for the NC-scale as large as  $10TeV$
the NC-effects on the CMB polarization and the r-parameter is
significant.  We show that the V-mode power spectrum can be
obtained in terms of linearly polarized power spectrum in the
range Micro to Nano-Kelvin squared for the NC-scale
about $1TeV$ to  $10TeV$, respectively.
\end{abstract}

\end{titlepage}
\section{Introduction}
The polarization anisotropy and temperature inhomogeneities of the cosmic microwave background radiation (CMBR) can provide a way for exploring the physics of the early universe.  The light polarization can be parameterized in terms of
the Stokes parameters ($Q, U$ and $V$). A  nonzero values for  $Q$ and/or $U$ show linearly polarized radiations while a circular polarized radiation has a non vanishing value for
 the Stokes parameter $V$\cite{jackson}.  An anisotropic Thomson scattering  due to the temperature inhomogeneity around the recombination phase leads to the linear polarization about $10\%$ of CMBR
\cite{kosowsky,Hu}.  Meanwhile, according to the standard model of cosmology there is no physical mechanism to
generate a circular polarized radiation at the last scattering
surface or $V=0$.  However, a linearly polarized radiation through its propagation in a magnetic field can be partially circular polarized,  a property known as the Faraday effect. The Stokes
parameter $V$ in this mechanism evolves as
\begin{equation}
\dot{V}=2\:U\frac{d\Delta\phi_{FC}}{dt},
\end{equation}
where $\Delta\phi_{FC}$ is the Faraday conversion phase shift \cite{Cooray:2002nm}.
Linear polarization of the CMB radiation from the last scattering can be converted to the circular polarization due to effects of background fields, particle scattering and temperature fluctuations.  The conversion probability of the CMB linear polarization to the circular polarization has been discussed in many papers \cite{gio1,gio2,cmbpol,Roh,xue}.  Furthermore, since  $Q$ and $U$ are frame-dependent, by decomposing the linear polarization into the $E$ and $B$  components one can extract more information from the polarization pattern on the sky. The Thomson scattering at the last scattering surface only produces the $E$ mode which can be converted to the $B$ mode through the vector and tensor perturbations.  Meanwhile, the gravitational waves, if exist,
due to the tensor mode  perturbation arising from the inflation epoch generate the B-mode polarization for
the CMB radiation.

 In early 2014, the BICEP2 team announced a nonzero measurement on the  B-mode polarization for
the CMB radiation as an evidence for the primordial gravitational wave\cite{BICEP2}. This result is not consistent with the Planck limit, $r<0.11 ( 98 \% CL)$. However, at this time there is no conclusive evidence of primordial gravitational waves from a joint analysis of data provided Planck and BICEP2 experiments. The recent
Bicep/Keck Array observation reported upper bounds on the tensor-to-scalar
ratio, $r_{0.05}<0.09$ and $r_{0.05}<0.07$ at $( 95\%)$ C.L. by using B-modes alone and combining the B-mode results
with Planck temperature analysis, respectively \cite{keck}. In fact, to distinguish the tensor and scalar components, a tensor-to-scalar ratio can be calculated by measuring the polarization angles on the sky, which Plank has been reported this ratio to be about $r\sim 0.12$.  Therefore, to find out the contribution of the gravitational wave on the $B$-mode one should consider all contribution from the other sources.   Although, in the standard model of cosmology the B-mode dose not receive contribution from the scalar mode one can consider the B-mode as a result of Faraday rotation of the E-mode polarization \cite{14,Giovannini}.  Furthermore,  the Compton
scattering in presence of a background can potentially lead to a B-mode polarization for the CMB even for the scalar perturbation.  The contribution to the observed B-mode spectrum from the interaction between CMB photon and the Cosmic Neutrino Background(CNB) in the scalar perturbation  background has been considered in \cite{roh}.

Here we would like to explore the effects of non-commutative background on the B-mode polarization.
  In ref.\cite{20Tev} the energy scale of the non-commutativity of space-time has been constrained by using CMB data from PLANCK.
They find that PLANCK data put the lower bound on the non-commutativity energy scale to about
20 TeV, which is about a factor of 2 larger than the previous bound that was obtained using data from WMAP, ACBAR and CBI.\\
In this paper we study the possibility of generating circular
polarization of  CMB radiation by considering Compton scattering
on Non-Commutative background. In Sec II we review the Stokes
parameters and Boltzmann  equation formalism. In Sec. III we give a
brief introduction on Non-Commutative standard model. In Sec. IV
the time evolution of Stokes parameters by using the scalar mode
perturbation of metric and the generation of circular polarization
on NonCommutative space is computed. Then We calculate circular,
E- and B- modes spectrum of CMB. By comparing our results with
experimental data the lower limit of non-commutative energy scale
is obtained.
\section{Stokes parameters and Boltzmann equation}
 For a monochromatic electromagnetic wave propagating in the $\hat{z}$ direction, the electric field components can be given as
\begin{eqnarray}
E_{x}=a_{x}\cos(\omega t-\theta_{x}), \hspace{2cm}  E_{y}=a_{x}\cos(\omega t-\theta_{y}),
\end{eqnarray}
where $a_{x}$ and $a_{y}$ are the amplitudes and $\theta_{x}$ and $\theta_{y}$ are the phase angles.
The electromagnetic field can be parameterized in terms of the Stokes parameters
 \begin{eqnarray}
I=\langle a_{x}^{2}\rangle+\langle a_{y}^{2}\rangle,
\end{eqnarray}
which is the total intensity and

 \begin{eqnarray}
Q=\langle a_{x}^{2}\rangle-\langle a_{y}^{2}\rangle,\hspace{1cm} ;  \hspace{1cm}    U=\langle 2a_{x}a_{y}\cos(\theta_{x}-\theta_{y})\rangle,
\end{eqnarray}
for the linear polarization.  $Q$ and $U$ are defined as the difference in brightness between the two linear polarization at $90^{o}$ and $45^{o}$, respectively, and the circular polarization is
 \begin{eqnarray}
V=\langle 2a_{x}a_{y}\sin(\theta_{x}-\theta_{y})\rangle.
\end{eqnarray}
 One can see that under a right handed rotation of the coordinate axes perpendicular to the direction $\hat{n}$ on the sky, $Q$ and $U$ with a rotation's angle $\psi$ transform to
\begin{eqnarray}
&&
Q'=Q\cos(2\psi)+U\sin(2\psi),
\nonumber
\\
&&
U'=-Q\sin(2\psi)+U\cos(2\psi),
\end{eqnarray}
and the Stokes parameters $I$ and $V$ remain unchanged.
 The density matrix in terms of the Stokes parameters is defined as
\begin{equation}
\rho=\frac{1}{2}
\left(
\begin{array}{cc}
I+Q & U-iV\\
U+iV & I-Q \\
\end{array}
\right).
\end{equation}
Meanwhile, a system of photons can be described by the density operator \cite{kosowsky}
\bea
\hat\rho=\frac{1}{\rm {tr}(\hat \rho)}\int\frac{d^3 \textbf k}{(2\pi)^3}
\rho_{ij}(\textbf k)a^\dagger_i(\textbf k)a_j(\textbf k),
\eea
where $\rho_{ij}(\textbf k)$ is the general density-matrix which is related to the photon number operator $D^0_{ij}(\textbf k)\equiv a_i^\dag (\textbf k)a_j(\textbf k)$.
The canonical commutation relations of the creation and annihilation operators for photons and electrons are defined as
\begin{eqnarray}\label{commutation}
 &&
[a_s(p),a_{s'}^\dagger(p')]=2\pi^32p^0\delta^3(\bold p-\bold p')\delta_{ss'},
\nonumber
\\
&&
\{b_r(q),b_{r'}^\dagger(q')\}=2\pi^3\frac{q^0}{m}\delta^3(\bold q-\bold q')\delta_{rr'},
\label{comm}
\end{eqnarray}
where $s$ and $r$ show the photon polarization and electron spin while the bold and plain momenta represent the three dimensional vectors and the four-momentum vectors, respectively.  For the expectation value of the number operator one has
\bea
\langle\, D^0_{ij}(\textbf k)\,\rangle\equiv {\rm tr}[\hat\rho
D^0_{ij}(\textbf k)]=(2\pi)^3 \delta^3(0)(2k^0)\rho_{ij}(\textbf k),\label{t1}
\eea
and in the Heisenberg picture, the time evolution of the operator $D^0_{ij}(\textbf k)$ can be obtained as
\begin{equation}
   \frac{d}{dt} D^0_{ij}(\textbf k)= i[H,D^0_{ij}(\textbf k)],\label{heisen}
\end{equation}
where $H$ is the total Hamiltonian. Therefore, the time evolution of the density-matrix can be written as follows
 \begin{eqnarray}\label{bo}
\hspace{-1cm}
(2\pi)^3 \delta^3(0)(2k^0)
\frac{d}{dt}\rho_{ij}(\textbf k)& =& i\langle \left[H^0_I
(t);D^0_{ij}(\textbf k)\right]\rangle  \\ \nonumber
& &-\frac{1}{2}\int dt\langle
\left[H^0_I(t);\left[H^0_I
(0);D^0_{ij}(\textbf k)\right]\right]\rangle,
\hspace{1cm}
 \end{eqnarray}
where $H^0_I(t)$ is the interacting
Hamiltonian at the lowest order. The first term on the right-hand side of (\ref{bo}) is a forward scattering term that is called
damping term, while the second term is the higher order collision term. To calculate the right hand side of  (\ref{bo})  one needs the operator expectation values as
\begin{eqnarray}
\langle a_1a_2...b_1b_2...\rangle=\langle a_1a_2...\rangle \langle b_1b_2...\rangle,
\end{eqnarray}
where for the two point functions  using the the commutation relations (\ref{commutation}) one has
\begin{eqnarray}
\langle a_{m}^\dagger(p')a_n(p)\rangle=2p^0(2\pi)^3\delta^3(\bold p-\bold p')\rho_{mn}(\bold p),
\end{eqnarray}
and
\begin{eqnarray}
&&
\langle b_{m}^\dagger(q')b_n(q)\rangle=\frac{q^0}{m}(2\pi)^3\delta^3(\bold q-\bold q')\delta_{mn}\frac{1}{2}n_f(\bold q),
\end{eqnarray}
where $n_f(\textbf q)$ represents the number density of fermions (electrons and  protons) with momentum $\textbf q$ per unit volume. Meanwhile, the distribution of fermions in the $x$ space are defined as
\begin{equation}\label{nd-n0}
n_f(\mathbf{x})=\int \frac{d^3 \bf{q}}{(2\pi)^3} \,\,n_f(\mathbf{x},\mathbf{q}),\quad m_fv_i(\mathbf{x})n_f(\mathbf{x})=\int\frac{ d^3\bf{q} }{(2\pi)^3}\, q_i\,n_f(\mathbf{x},\mathbf{q}).
\end{equation}
\section{Noncommutative standard model}

The noncommutative space-time is one of the consequence of the string theory.
Indeed, as is predicted in \cite{ardalan} and shown by Seiberg and
Witten \cite{seiberg}, endpoints on D-branes  in a constant B-field
background live on a noncommutative space-time. In the canonical form one has
\begin{equation}
\theta^{\mu\nu}=-i\left[\hat{x}^\mu,\hat{x}^\nu\right],
\end{equation}
where $\theta^{\mu\nu}$ is a real, constant and  antisymmetric
matrix; $\theta^{\mu \nu} \propto 1/\Lambda^2_{\tiny{NC}}$, and
$\Lambda_{\tiny{NC}}$ is the noncommutative scale. The noncommutative parameter, $\theta^{\mu\nu}$, can be divided into two parts:
the time-space components $(\theta^{01},\theta^{02},\theta^{03})$ which denote the electric-like part and the magnetic-like
part which contains the space-space components
$(\theta^{23},\theta^{31},\theta^{12})$. To construct  noncommutative filed theories, one should replace
the ordinary products between fields in the corresponding commutative versions with the Moyal-$\star$
product which is defined as follows
\begin{equation}
(f\star
g)(x)=f(x)\exp(\frac{i}{2}\overleftarrow{\partial_\mu}
\theta^{\mu\nu}\overrightarrow {\partial_\nu})g(x).
\label{a3}
\end{equation}
By applying this correspondence, there would be two approaches to
construct the noncommutative standard model. The first one is based on the Moyal-$\star$ product and the Seiberg-Witten maps in which the gauge group is $SU(3)\times SU(2)_L\times U(1)_Y$, the number of
particles and gauge fields are the same as the ordinary standard
model\cite{Calmet}.  Furthermore, matter fields, gauge fields and gauge parameters should be expanded via the Seiberg-Witten map as a power series of $\theta$\cite{Calmet} in terms of the commutative fields as

 \begin{eqnarray}
\widehat{\psi}&=&\psi+\frac{1}{2}\theta_{\mu\nu}A_{\nu}\partial_{\mu}\psi+\mathcal{O}(\theta^2), \\ \nonumber
 \widehat{A}_{\mu}&=&A_{\mu}+\frac{1}{4}\theta^{\rho\nu}\{A_{\nu},(\partial_{\rho}A_{\mu}+F_{\rho\mu})\}+\mathcal{O}(\theta^2),\\ \nonumber
\widehat{\Lambda}&=&\Lambda+\frac{1}{4}\theta^{\mu\nu}\{A_{\nu},\partial_{\mu}\Lambda\}+\mathcal{O}(\theta^2),
\label{gaie}
 \end{eqnarray}
 Where the hats show the noncommutative fields which reduce to  their counterparts in the ordinary space in the limit $\theta \rightarrow 0$.\\
 In the second approach the noncommutative fields and the ordinary fields are the same while the gauge group is  $U(3)\times U(2)\times U(1)$ which is reduced
 to $SU(3)\times SU(2)_L\times U(1)_Y$ by an appropriate symmetry breaking\cite{ncf2}.
 In the both versions, besides corrections on the usual standard model interactions, many new interactions would appear.  For instance, in the QED part of the NCSM through the first approach there is a correction on the photon-fermion vertex $ff\gamma$ that can be derived up to the first order of $\theta$ as \cite{melic}
\\
\\
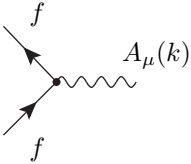
\begin{picture}(55,45) (30,-30)
\SetWidth{0.5}
\ArrowLine(30,-30)(50,-10)
\ArrowLine(50,-10)(30,11)
\Photon(80,-10)(50,-10){2}{4}
\Vertex(50,-10){1.5}
\Text(40,11)[lb]{$f$}
\Text(40,-40)[lb]{$f$}
\Text(75,-5)[lb]{$A_{\mu}(k)$}
\end{picture}
\\
\begin{eqnarray}
\lefteqn{i \, e \, Q_{f}\,
  \left[
 \gamma_{\mu}
 - \frac{i}{2} \, k^{\nu}
   \left( \theta_{\mu \nu \rho}
 \, p_{\mbox{\tiny in}}^{\rho}-
\theta_{\mu \nu}\, m_f\,\right)
  \right]}
\nonumber \\[0.2cm]
& = &
i \, e \, Q_{f}\,
\gamma_{\mu}
\nonumber \\
& &
+ \frac{1}{2} \, e \, Q_{f}\,
\left[
(p_{\mbox{\tiny out}} \theta p_{\mbox{\tiny in}}) \gamma_\mu
-
(p_{\mbox{\tiny out}}\theta)_\mu(\slash \!\!\! p_{\mbox{\tiny in}}-m_f)
-
(\slash \! \! \! p_{\mbox{\tiny out}}-m_f)(\theta p_{\mbox{\tiny in}})_\mu
\right]\,.
\nonumber \\
\label{eq:ffgamma}
\end{eqnarray}
Meanwhile, in the noncommutative QED there are vertices which have not any counterpart in the ordinary QED. For example, two photons can directly couple to two fermions in NC space as follows
\\
\\
\\
\\
\\
\begin{picture}(55,45) (30,-30)
\SetWidth{0.5}
\ArrowLine(30,-30)(50,-10)
\ArrowLine(50,-10)(30,11)
\Photon(70,-30)(50,-10){2}{4}
\Photon(70,11)(50,-10){2}{4}
\Vertex(50,-10){1.5}
\Text(40,11)[lb]{$f$}
\Text(40,-40)[lb]{$f$}
\Text(77,-40)[lb]{$A_{\nu}(p)$}
\Text(77,8)[lb]{$A_{\mu}(p')$}
\end{picture}
\vspace{-1.5cm}
\begin{eqnarray}
\frac{-\,e^2\, Q^2_f}{2} \,
    \theta_{\mu\nu\rho} \, (p'^{\rho}-p^{\rho}) \,,
\label{eq:ffgammagamma}
\end{eqnarray}\\
\\
\\
where photons momenta are taken to be incoming and   $\theta^{\mu
\nu \rho}$ is a totally antisymmetric quantity which is defined as
\begin{eqnarray}
\theta^{\mu \nu \rho}=
\theta^{\mu \nu} \gamma^{\rho}
+ \theta^{\nu \rho} \gamma^{\mu}
+ \theta^{\rho \mu} \gamma^{\nu}\,.
\label{eq:theta3}
\end{eqnarray}

\section{Compton scattering in NC space-time}
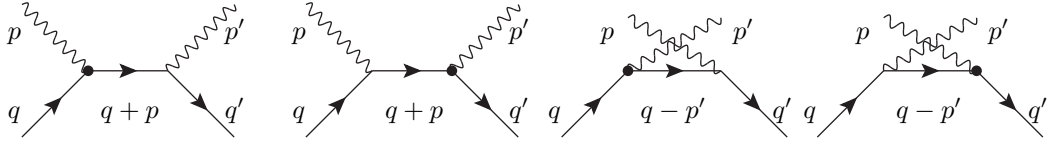
\begin{figure}
\begin{picture}(55,45) (30,-30)
\SetWidth{0.5}
\hspace{-.5cm}
\Photon(50,-25)(75,-50){2}{7}
\ArrowLine(50,-75)(75,-50)
\Vertex(75,-50){2.0}
\ArrowLine(75,-50)(105,-50)
\Photon(105,-50)(130,-25){2}{7}
\ArrowLine(105,-50)(130,-75)
\Text(45,-70)[lb]{$q$}
\Text(45,-40)[lb]{$p$}
\Text(80,-70)[lb]{$q+p$}
\Text(127,-40)[lb]{$p'$}
\Text(127,-70)[lb]{$q'$}
\end{picture}
\hspace*{1.25cm}
\begin{picture}(55,45) (30,-30)
\SetWidth{0.5}
\Photon(50,-25)(75,-50){2}{7}
\ArrowLine(50,-75)(75,-50)
\Vertex(105,-50){2.0}
\ArrowLine(75,-50)(105,-50)
\Photon(105,-50)(130,-25){2}{7}
\ArrowLine(105,-50)(130,-75)
\Text(45,-70)[lb]{$q$}
\Text(45,-40)[lb]{$p$}
\Text(80,-70)[lb]{$q+p$}
\Text(127,-40)[lb]{$p'$}
\Text(127,-70)[lb]{$q'$}
\end{picture}
\hspace*{1.25cm}
\begin{picture}(55,45) (30,-30)
\SetWidth{0.5}
\Photon(110,-30)(75,-50){2}{7}
\ArrowLine(50,-75)(75,-50)
\Vertex(75,-50){2.0}
\ArrowLine(75,-50)(110,-50)
\Photon(110,-50)(75,-30){2}{7}
\ArrowLine(110,-50)(135,-75)
\Text(45,-70)[lb]{$q$}
\Text(65,-40)[lb]{$p$}
\Text(80,-70)[lb]{$q-p'$}
\Text(115,-40)[lb]{$p'$}
\Text(130,-70)[lb]{$q'$}
\end{picture}
\hspace*{1.25cm}
\begin{picture}(55,45) (30,-30)
\SetWidth{0.5}
\Photon(110,-30)(75,-50){2}{7}
\ArrowLine(50,-75)(75,-50)
\Vertex(110,-50){2.0}
\ArrowLine(75,-50)(110,-50)
\Photon(110,-50)(75,-30){2}{7}
\ArrowLine(110,-50)(135,-75)
\Text(45,-70)[lb]{$q$}
\Text(65,-40)[lb]{$p$}
\Text(80,-70)[lb]{$q-p'$}
\Text(115,-40)[lb]{$p'$}
\Text(130,-70)[lb]{$q'$}
\end{picture}
\vspace*{1.75cm}
\caption{Compton scattering in NC space}
\label{Compton}
\end{figure}

The photons in the cosmic Microwave background can be scattered from all charged
particles with the scattering rate proportional to the
inverse mass squared.  Therefore, in the ordinary space as a good approximation the Compton scattering on the
electrons is usually considered. There are 5 diagrams in the NCSM to describe the Compton scattering on noncommutative
space-time.  By replacing the ordinary couplings with the NC vertices, four diagrams can be obtained which is shown in  Fig \ref{Compton}.  The fifth diagram in which two fermions directly couple to two photons is given in (\ref{eq:ffgammagamma}).  Therefore, the amplitude of the Compton scattering in the NCQED can be given as
\bea
\mathcal{M_{NCQED}}=\mathcal{M_{QED}}+\mathcal{M}^{\theta},
\label{mnc}
\eea
where
\bea
i \mathcal{M}^{\theta}\!\!\!\!\!\!\!\!&&=i\mathcal{M}^{\theta}_1+i\mathcal{M}^{\theta}_2+i\mathcal{M}^{\theta}_3
+i\mathcal{M}^{\theta}_4+i\mathcal{M}^{\theta}_5 \nonumber\\&&=\frac{\:e^2Q_f^2}{4p\cdot
q}\bar{u}_{\acute{r}}(q')\Big[\Big(\ep_{\acute{s}}\!\!\!\!\!\!/\:\:(\acute{p})\acute{q}\cdot\theta\cdot
(q+p)-\acute{q}\cdot\theta\cdot\ep_{\acute{s}}(\acute{p})\acute{p}\!\!\!\!/\:\Big)(q\!\!\!/+p\!\!\!/+m_f)\ep_{s}\!\!\!\!\!/\:\:(p)\no\\&&+
\ep_{\acute{s}}\!\!\!\!\!\!/\:\:(\acute{p})(q\!\!\!/+p\!\!\!/+m_f)\Big(\ep_{s}\!\!\!\!\!/\:\:(p)(q+p)\cdot\theta\cdot
q -q\cdot\theta\cdot\ep_{s}(p)p\!\!\!/\:\Big)\Big]u_{r}(q)\no\\&&
-\frac{\:e^2Q_f^2}{4\acute{p}\cdot
q}\bar{u}_{\acute{r}}(\acute{q})\Big[\Big(\ep_{s}\!\!\!\!\!/\:\:(p)\acute{q}\cdot\theta\cdot
(q-\acute{p})+\acute{q}\cdot\theta\cdot\ep_{s}(p)p\!\!\!/\:\Big)(q\!\!\!/-p'\!\!\!\!/+m)\ep_{\acute{s}}\!\!\!\!\!\!/\:\:(\acute{p})\no\\&&+
\ep_{s}\!\!\!\!\!/\:\:(p)(q\!\!\!/-\acute{p}\!\!\!\!/+m_f)\Big(\ep_{\acute{s}}\!\!\!\!\!\!/\:\:(\acute{p})(q-\acute{p})\cdot\theta\cdot
q +q\cdot\theta\cdot\ep_{\acute{s}}(\acute{p})\acute{p}\!\!\!/\:\Big)\Big]u_{r}(q)\nonumber\\&&
-\frac{e^2Q_f^2}{2} \bar{u}_{\acute{r}}(q') \ep_{\acute{s}}^\mu (p') (p+p')^\rho \theta_{\mu\nu\rho} \ep_s ^\nu (p)  u_r(q),
\label{m} \eea
and $p\cdot\theta\cdot q\equiv p_{\mu}\theta^{\mu\nu}q_{\nu}$.  Now the leading-order interacting Hamiltonian can be obtained as follows
\begin{eqnarray}
  H^0_I &=& \int d\mathbf{q} d\mathbf{q'} d\mathbf{p} d\mathbf{p'} (2\pi)^3\delta^3(\mathbf{q'} +\mathbf{p'} -\mathbf{p} -\mathbf{q} ) \nonumber \\
   &\times& \exp[it(q'^0+p'^0-q^0-p^0)]\left[b^\dagger_{r'}a^{\dagger}_{s'}(\mathcal{M^{\theta}})a_sb_r\right],\label{h0}
   \label{H}
\end{eqnarray}
where $d\mathbf{q}\equiv \frac{d^3q}{(2\pi)^3}\frac{m_f}{q^0}$ and $d\mathbf{p}\equiv \frac{d^3p}{(2\pi)^32p^0}$, and the similar expressions for $d\mathbf{p'}$ and $d\mathbf{q'}$, respectively.\\

\section{Density matrix elements and CMB Polarization on NC space-time}
In the preceding section we found the amplitude for the Compton scattering in the NCQED.  Now substituting the NC interacting Hamiltonian  ($\ref{H}$) in ($\ref{bo}$) leads to
 \bea
i\left<[\mathcal{\hat H}_{int}(0),\mathcal{D}_{ij}(\textbf{k})]\right>\!\!\!\!\!\!\!&&=-\frac{e^2Q_f^2}{2}\:\delta^{3}(0)\int
d\textbf{q}\:(2\pi)^3 n_f(\textbf{x},\textbf{q})(\d_{is}\rho_{s'j}(\textbf{k})-\d_{js'}\rho_{is}(\textbf{k}))\no\\&&
\!\!\!\!\!\!\!\!\!\!\!\!\!\!\!\!\!\!\!\!\!\!\!\!\!\!\!\!\!\!\!\!\!\!\!\!\!\!\!\!\!\!\!\!\!\!\!
 \times
 \bar{u}_{r}(q)\{ \frac{1}{4 k\cdot q} \Big[q.\theta.\ep_{s'}(k)\Big(k\!\!\!\!/\:\:(q\!\!\!/+k\!\!\!/+m_f)\ep_{s}(k)\!\!\!\!\!/\:\:
 +\ep_{s}\!\!\!\!\!/\:\:(k)(q\!\!\!/-k\!\!\!/+m_f)
 k\!\!\!/\Big)\no\\&&
\hspace{-3cm}+q.\theta.\ep_{s}(k)\Big(\ep_{s'}\!\!\!\!\!/\:\:(k)(q\!\!\!/+k\!\!\!/+m_f)
 k\!\!\!/+k\!\!\!\!/\:\:(q\!\!\!/-k\!\!\!/+m_f)\ep_{s'}(k)\!\!\!\!\!/\:\:\Big)
\Big] \no\\&& \hspace{-3cm} + \Big[\ep_{s'}(k)\cdot
\theta\cdot \ep_s(k)\:\: k\!\!\!\!/\:\:- k \cdot\theta \cdot
\epsilon_s(k) \:\:\ep_{s'}\!\!\!\!\!/\:\:(k)\!\!\!\!\!\:\:\: + k
\cdot\theta \cdot \ep_{s'}(k)
\:\:\ep_{s}\!\!\!\!\!/\:\:(k)\Big] \}u_{r}(q). \hspace{1cm}\eea
Therefore,  the time evaluation of the density matrix after a little algebra can be obtained as
 \bea
\frac{d}{dt}{\rho}_{ij}(\textbf{k})=-\frac{e^2Q_f^2}{4m_fk^0}\int d\textbf{q}
\:n_f(\textbf{x},\textbf{q})(\delta_{is}
\rho_{\acute{s}j}-\delta_{j\acute{s}}
\rho_{is})\Big(A+B\Big),
~~~~~~~~\label{rho01} \eea
where
\bea
A=q\cdot\theta\cdot\ep_{\acute{s}}~q\cdot\ep_{s}+q\cdot\theta\cdot\ep_{s}~q\cdot\ep_{\acute{s}},\label{A}
\eea
and
\bea
B= \Big(\ep_{s'}(k)\cdot \theta\cdot
\ep_s(k)\:\: k\cdot q - k \cdot \theta\cdot
\ep_{s}(k)\:\: \ep_{s'}(k).q + k\cdot \theta\cdot
\ep_{s'}(k)\:\: \ep_{s}(k).q \Big).\label{B}
\eea
It should be noted that $A$ and $B$ are related to the NC contribution on
the time evaluation of the density matrix from the vertices given in  (\ref{eq:ffgamma} ) and
(\ref{eq:ffgammagamma}), respectively.  As we show below the first contribution depends on the time-like component of the NC
parameter $\theta^{0i}$ while the second contribution depends on the
 space-like component of the NC parameter
$\theta^{ij}$.  To this end we individually evaluate the contribution of each part on the  time evolution of the stokes parameters as follows\\

{\bf{Part A:}}

For the $A$ term of (\ref{rho01}) which is due to the diagrams given in Fig.(\ref{Compton}) the time evolution of the density matrix is
\bea
\frac{d}{dt}{\rho}_{ij}(\textbf{k})\!\!\!\!\!\!&&=
-\frac{e^2Q_f^2}{4m_fk^0}\int d\textbf{q}
\:n_f(\textbf{x},\textbf{q})(\delta_{is} \rho_{\acute{s}j}-\delta_{j\acute{s}} \rho_{is})A.
~~~~~~~~\label{rho1}
\eea
By using Eq.(\ref{nd-n0}) and assuming
\bea
\theta^{01}=\theta^{02}=\theta^{03}=\frac{1}{\Lambda^2},\,\,\,\,\,\hat{\theta}^{0i}=\Lambda^2\theta^{0i},\label{ass1}
\eea
and
\bea
\theta^{ij}=\frac{1}{\Lambda^2},\,\,\,\,\,\hat{\theta}^{ij}=\Lambda^2\theta^{ij},\label{ass1}
\eea
where $\theta^{ij}$ ; $i,j\in \{1,2,3\}$ are the space-like components of the NC-parameter, one can rewrite (\ref{rho1}) as follows
\bea
\frac{d}{dt}{\rho}_{ij}(\textbf{k})\!\!\!\!\!\!&=&
\frac{e^2Q_f^2}{4}\frac{m_f}{k^0}\frac{1}{\Lambda^2}
\:\bar{n}_f(\textbf{x})(\delta_{is} \rho_{\acute{s}j}-\delta_{j\acute{s}} \rho_{is})\nonumber\\
&\times&\Big[
\,\Big(\hat{\theta}^{0i}\ep_{\acute{s}i}~v_f\cdot\ep_{s}+\hat{\theta}^{0i}\ep_{si}~v_f\cdot\ep_{\acute{s}}\Big)\nonumber\\
&+&\frac{1}{\sqrt{2}}\Big(\ep_{s}\cdot\theta\cdot\ep_{\acute{s}}+\ep_{\acute{s}}\cdot\theta\cdot\ep_{s}\Big)
+\bigcirc(v_f^2)\Big],~~~~~~~~\label{rho2} \eea
 where the polarization four-vectors $\ep_{\mu i}(k)$ with $i, j, s$ and $s'$ run over $1,
2$, represent two transverse polarization of photon,
$\bar{n}_f$ represents the number density of Fermions  and $v_f$
is the Fermion bulk velocity which is a small quantity. Note that the first term in the third line of (\ref{rho2}) vanishes due to the anti-symmetric property of
$\theta$ and the second one depends on $v_f^2$ which is negligible in comparison with the first term. Therefore, the time
derivative of the components of the density matrix can be cast into
 \bea
\frac{d}{dt}\rho_{11}(\textbf{k})\!\!\!\!\!\!&&=
\frac{3}{8}\frac{m_f}{k^0}\frac{\sigma^T}{\alpha}\,\frac{m_e^2}{\Lambda^2}\,\bar{n}_f\,\hat{\theta}^{0i}\,
\Big(\,\ep_{2i}~v_f\cdot\ep_1+\ep_{1i}~v_f\cdot\ep_2\Big)(\rho_{21}-\rho_{12}),\label{rho11}\\
\frac{d}{dt}\rho_{22}(\textbf{k})\!\!\!\!\!\!&&=\frac{3}{8}\frac{m_f}{k^0}\frac{\sigma^T}{\alpha}\,\frac{m_e^2}{\Lambda^2}\,\bar{n}_f\,\hat{\theta}^{0i}
\Big(\ep_{2i}~v_f\cdot\ep_1+\ep_{1i}~v_f\cdot\ep_2\Big)(\rho_{12}-\rho_{21}),\label{rho22}\\
\frac{d}{dt}\rho_{12}(\textbf{k})\!\!\!\!\!\!&&=
\frac{3}{8}\frac{m_f}{k^0}\frac{\sigma^T}{\alpha}\,\frac{m_e^2}{\Lambda^2}\,\bar{n}_f\,\hat{\theta}^{0i}\,
\Big(2(\ep_{1i}~v_f\cdot\ep_1-\ep_{2i}~v_f\cdot\ep_2)\rho_{12}
\no\\&& \hspace{-1.5cm}
-(\ep_{2i}~v_f\cdot\ep_{1}+\ep_{1i}~v_f\cdot\ep_2)\rho_{11}
+(\ep_{2i}~v_f\cdot\ep_1
+\ep_{1i}~v_f\cdot\ep_2)\rho_{22}\Big)\label{ro1},\hspace{1cm},\label{rho12}\\
\frac{d}{dt}\rho_{21}(\textbf{k})\!\!\!\!\!\!&&=-\frac{3}{8}\frac{m_f}{k^0}\frac{\sigma^T}{\alpha}\,\frac{m_e^2}{\Lambda^2}
\,\bar{n}_f\,\hat{\theta}^{0i}\,
\Big(2(\ep_{1i}~v_f\cdot\ep_1-\ep_{2i}~v_f\cdot\ep_2)\rho_{21}\no\\&&
\hspace{-1.5cm}
-(\ep_{2i}~v_f\cdot\ep_1+\ep_{1i}~v_f\cdot\ep_2)\rho_{11}
+(\ep_{2i}~v_f\cdot\ep_1
+\ep_{1i}~v_f\cdot\ep_2)\rho_{22}\Big)\label{ro2},
\hspace{1cm}\label{rho21}\eea
where $m_e$ is the mass of electron,
$\sigma^T$ is the Thomson cross section, $\alpha=e^2/4\pi$ and
$Q_f^2=1$. Using the density operator matrix elements, time
variation of the stokes parameters, linear  polarization intensities $Q$ and $U$ and the difference between the left and right handed polarization $V$ in the NC space can be obtained as follows
\begin{eqnarray}
&&
\dot{V}(\textbf{k})=i\frac{3}{4}\frac{m_f}{k^0}\frac{\sigma^T}{\alpha}\,\frac{m_e^2}{\Lambda^2}\,\bar{n}_f(CQ+DU),\label{V}
\\
&&
\dot{Q}(\textbf{k})=i\frac{3}{4}\frac{m_f}{k^0}\frac{\sigma^T}{\alpha}\,\frac{m_e^2}{\Lambda^2}\,\bar{n}_f(-CV),\label{Q}
\\
&&
\dot{U}(\textbf{k})=i\frac{3}{4}\frac{m_f}{k^0}\frac{\sigma^T}{\alpha}\,\frac{m_e^2}{\Lambda^2}\,\bar{n}_f(-DV);\label{U}
\end{eqnarray}
where
\begin{eqnarray}
&&
C = -\hat\theta^{0i}\Big(\ep_{1i}~v_f\cdot\ep_{2} + \ep_{2i}~v_f\cdot\ep_{1}\Big),
\nonumber
\\
&&
D = \hat\theta^{0i}\Big(\ep_{1i}~v_f\cdot\ep_{1} - \ep_{2i}~v_f\cdot\ep_{2}\Big) .\label{CD}
\end{eqnarray}
These equations show that the contribution of the $A$ term on the time evolution of the stokes parameters depends on the mass and
bulk velocity of Fermion and the time-like components of the NC-parameter $\theta^{0i}$ and as is already claimed there is not any  contribution from the space-space part of the NC-parameter. In contrast with the usual Compton
scattering which has a larger cross section for the particles with lower masses the evolution of the stokes parameters in the NC space are directly proportional to the Fermion masses which leads to the larger values for the
scattering from Fermions with larger masses, see (\ref{V}-\ref{U}).  In fact, since
the average number of electrons
$\bar{n}_e$ approximately equals to the average number of protons $\bar{n}_p$ due to electric neutrality in cosmology, in the NC space-time the contribution of photon-proton forward scattering  is larger
than photon-electron scattering on the evolution of the Stokes parameters by a factor $m_p/m_e$.
 Nevertheless,
 to have any significant effects from the $A$ term on the CMB polarization, the factor
$\frac{3}{8}\frac{m_f}{k^0}\frac{1}{\alpha}\,\frac{m_e^2}{\Lambda^2}$ should be comparable to one. It should be noted that $\frac{m_f}{k^0}$ is much larger than unity which can compensate the smallness of $\frac{m_e^2}{\Lambda^2}$.

{\bf Part B:}

For the $B$ term of (\ref{rho01}) which is coming from the direct vertex (\ref{eq:ffgammagamma}) we have
\bea
\frac{d}{dt}{\rho}_{11}(\textbf{k})\!\!\!\!\!\!&&= - \frac{d}{dt}{\rho}_{22}(\textbf{k})=
-\frac{e^2}{4m_f k^0}\int d\textbf{q}
\:n_f(\textbf{x},\textbf{q})~ (\rho_{12}(\textbf{k})+\rho_{21}(\textbf{k}))\times
\no\\&&
\Big[\ep_{2}(k)\cdot \theta\cdot \ep_1(k)\:\: k\cdot q - k \cdot \theta\cdot \ep_{1}(k)\:\: \ep_{2}(k)\cdot q
+ k\cdot \theta\cdot \ep_{2}(k)\:\: \ep_{1}(k)\cdot q
\Big],\no\\
\frac{d}{dt}{\rho}_{12}(\textbf{k})\!\!\!\!\!\!&& = \frac{d}{dt}{\rho}_{21}(\textbf{k})=
-\frac{e^2}{4m_fk^0}\int d\textbf{q}
\:n_f(\textbf{x},\textbf{q}) ~(\rho_{22}(\textbf{k})-\rho_{11}(\textbf{k})),\label{SL-rho}\\&&
\Big[\ep_{2}(k)\cdot \theta\cdot \ep_1(k)\:\: k\cdot q - k \cdot \theta\cdot \ep_{1}(k)\:\: \ep_{2}(k)\cdot q
+ k\cdot \theta\cdot \ep_{2}(k)\:\: \ep_{1}(k)\cdot q
\Big],\no
\eea

which after some calculations, lead to
\bea
\dot{I}(\textbf{k})&=&0,~~~~~~~~~~~\dot{V}(\textbf{k})=0,\no\\
\dot{Q}(\textbf{k}) \pm i~\dot{U}(\textbf{k}))&=& \pm ~ 2~i~F~(Q(\textbf{k}) \pm i~U(\textbf{k})),
\eea
where
\bea
F &=& \sigma|_{_{NC dv}}\,\bar{n}_f\,\,
\Big\{\hat{\theta}^{0i} \Big(\,\ep_{1i}~v_f\cdot\ep_2-\ep_{2i}~v_f\cdot\ep_1\Big)
\no\\
&+& \hat{\theta}^{ij}\big[\ep_{2}\cdot \theta\cdot \ep_1(k)\:\: (1 - v_f \cdot \hat{k}) + \hat{k} \cdot \theta\cdot \ep_{1}\:\: v_f\cdot \ep_{2}
- \hat{k}\cdot \theta\cdot \ep_{2}\:\: v_f\cdot \ep_{1} \big]\Big\},
\eea
and
\bea
\sigma|_{_{NC dv}}=\frac{3}{8}\frac{\sigma_T}{\alpha}\frac{m_e^2}{\Lambda^2}.
\eea
Here the time evolution of the stokes parameters depend on the space-space part of the NC-parameter as well.  However, in order to have any significant effect on the CMB polarization, the value of $\sigma|_{_{NC dv}}$ should be comparable to $\sigma_{T}$.  For the photon-proton scattering one has
\begin{equation}\label{dirct-vertex1}
    \sigma|_{_{NC dv}}/\sigma_{T}\propto\frac{1}{\alpha}(\frac{m_e}{\Lambda})^2< 10^{-10}(\frac{TeV}{\Lambda})^2,
\end{equation}
which is too small to be considered. Therefore, we can neglect the $B$ term with respect to the $A$ term to evaluating the CMB
polarization in the next sections.
\section{Time-evolution of polarized CMB photons}
In this section we expand the primordial scalar perturbations $(S)$ in the Fourier modes which is characterized by a wave-number $\mathbf{K}$. For a given
Fourier mode $\mathbf{K}$, one can select a coordinate system
where $\mathbf{K} \parallel \hat{\mathbf{z}}$ and
$(\hat{\mathbf{e}}_1,\hat{\mathbf{e}}_2)=(\hat{\mathbf{e}}_\theta,
\hat{\mathbf{e}}_\phi)$. We consider the  electron and baryon
bulk velocity directions as $\vec{v}_e =\vec{v}_b \parallel
\mathbf{K}$ and the photon polarization vectors are taken to be
\begin{eqnarray}
&&
\hat{\ep}_{1x}=\cos{\theta}\cos{\varphi},
\hspace{0.2cm}
\hat{\ep}_{1y}=\cos{\theta}\sin{\varphi},
\hspace{0.2cm}
\hat{\ep}_{1z}=-\sin{\theta},
\nonumber
\\
&&
\hat{\ep}_{2x}=-\sin{\varphi},
\hspace{0.2cm}
\hat{\ep}_{2y}=\cos{\varphi},
\hspace{0.2cm}
\hat{\ep}_{2z}=0.\label{condition}
\end{eqnarray}

Meanwhile, temperature anisotropy (I) and
polarization (Q,U) of the CMB radiation can be expanded in an
appropriate spin-weighted basis as follows\cite{zal}
\begin{eqnarray}
&&
\Delta^{(S)}_{I}(\mathbf{K},\mathbf{k},\tau)=\sum_{\ell m}a_{\ell m}(\tau,K)Y_{lm}(\mathbf{n}),\label{AA0}\\
&&
\Delta^{\pm (S)}_{P}(\mathbf{K},\mathbf{k},\tau)=\sum_{\ell m}a_{\pm2,\ell m}(\tau,K) _{\pm2}Y_{lm}(\mathbf{n}),\label{AA}
\end{eqnarray}
and
\bea
\Delta^{(S)}_I(\mathbf{K},\mathbf{k},\tau)=\left(4k\frac{\partial I_0}{\partial k}\right)^{-1} \Delta^{(S)}_ I(\mathbf{K},\mathbf{k},t),\,\,\,\,\,\,\,\Delta _{P}^{\pm (S)}=Q^{(S)}\pm iU^{(S)}.
\eea
For each plane wave, the scattering can be described as the
transport through a plane parallel medium \cite{chandra,kaiser}, which leads to the
Boltzmann equations as
\begin{eqnarray}
&&\frac{d}{d\eta}\Delta_I^{(S)} +iK\mu \Delta_I^{(S)}+4[\dot{\psi}-iK\mu \varphi]
=C^I_{e\gamma} , \label{Boltzmann}\\
&&\frac{d}{d\eta}\Delta _{P}^{\pm (S)} +iK\mu \Delta _{P}^{\pm (S)} = C^\pm_{e\gamma}- \, iv_b\kappa_{NC}^{\pm}\,\Delta _{V}^{(S)},\label{Boltzmann0}
\end{eqnarray}
and
\bea
\frac{d}{d\eta}\Delta _{V}^{(S)} +iK\mu \Delta _{V}^{(S)} = C^V_{e\gamma}+ i\frac{v_b}{2}\Big[\kappa_{NC}^{-}\,\Delta _{P}^{+(S)}+\kappa_{NC}^{+}\,\Delta _{P}^{-(S)}\Big],
\label{Boltzmann1}
\eea
here $C^I_{e\gamma}$, $C^\pm_{e\gamma}$ and $C^V_{e\gamma}$ indicate the contributions from the usual
photon-electron Compton scattering to the time evaluation of $I$,
$\Delta _{P}^{\pm (S)}$ and $V$ parameters,
respectively, their expressions can
be found for example in
\cite{kosowsky,zal,hu}.  In (\ref{Boltzmann}-\ref{Boltzmann1})  $\mu=\hat{\bf n}\cdot
\hat{\mathbf{K}}=\cos\theta$, $\theta$ is the angle between the direction of the CMB photon  $\hat{\bf n}={\bf k}/|{\bf k}|$ and the wave-vectors $\mathbf{K}$ and
\begin{equation}\label{kappa}
    \kappa_{NC}=a(\eta)\frac{3}{4}\frac{\sigma^T}{\alpha}\,\frac{m_e^2}{\Lambda^2}\,\bar{n}_e\sum_{f=e,p}\frac{m_f}{k^0},\,\,\,\,\,\,
    \kappa_{NC}^{\pm}=\kappa_{NC}(C\pm iD),
\end{equation}
where $C$ and $D$ are given in (\ref{CD}) and  $a(\eta)$ is the normalized scaling factor. The values of
$\Delta _{P}^{\pm (S)}(\hat{n})$ and $\Delta _{V}^{(S)}$  at
the present time $\tau_0$ and the direction $\hat{n}$ are obtained  by integrating the Boltzmann equation (\ref{Boltzmann1}) along the line of sight \cite{zal} and summing over all the Fourier modes $K$ as follows
\begin{eqnarray}
\Delta _{P}^{\pm (S)}(\hat{\bf{n}})
&=&\int d^3 \bf{K} \xi(\bf{K})e^{\pm2i\phi_{K,n}}\Delta _{P}^{\pm (S)}
(\mathbf{K},\mathbf{k},\tau),\,\,\,\,\,\label{Boltzmann03}\\
\Delta _{V}^{ (S)}(\hat{\bf{n}})
&=&\int d^3 \bf{K} \xi(\bf{K})\Delta _{V}^{(S)}
(\mathbf{K},\mathbf{k},\tau),\,\,\,\,\,\label{Boltzmann3}
\eea
where $\phi_{K,n}$ is the angle needed to
rotate the $\bf{K}$ and $\hat{\bf{n}}$ dependent basis to a fixed
frame in the sky, $\xi(\bf{K})$ is a random variable using to
characterize the initial amplitude of the mode, and
\bea
 \Delta _{P}^{\pm (S)}
(\mathbf{K},\mu,\tau_0)&=&\int_0^{\tau_0} d\tau\,\dot\tau_{e\gamma}\,
e^{ix \mu -\tau_{e\gamma}}\,\,\Big[ {3 \over 4}(1-\mu^2)\Pi(K,\tau)\nonumber\\
&-& iv_b\frac{\kappa_{NC}^{\pm}}{\dot\tau_{e\gamma}}\,\Delta _{V}^{(S)}\Big],
\end{eqnarray}
and
\bea
 \Delta _{V}^{(S)}
(\mathbf{K},\mu,\tau_0)&=&\frac{1}{2}\int_0^{\tau_0} d\tau\,
\dot\tau_{e\gamma}\,e^{ix \mu -\tau_{e\gamma}}\,\,\Big[ 3\mu\Delta _{V1}^{(S)}\nonumber\\
&+&iv_b(\frac{\kappa_{NC}^{-}}{\dot\tau_{e\gamma}}\,\Delta _{P}^{+(S)}+\frac{\kappa_{NC}^{+}}{\dot\tau_{e\gamma}}\,\Delta _{P}^{-(S)})\Big],\nonumber\\
&\approx&\frac{1}{2}\int_0^{\tau_0} d\tau\,
\dot\tau_{e\gamma}\,e^{ix \mu -\tau_{e\gamma}}\,\,\Big[ 3\mu\Delta _{V1}^{(S)}+2iv_b\,C\,\frac{\kappa_{NC}}{\dot\tau_{e\gamma}}\,\Delta _{P}^{(S)}\Big],
\eea
in which $x=K(\tau_0 - \tau)$, $C$ is defined in (\ref{CD}) and
\begin{equation}\label{DP}
   \Delta _{P}^{(S)}
(\mathbf{K},\mu,\tau)=\int_0^{\tau} d\tau\,\dot\tau_{e\gamma}\,
e^{ix \mu -\tau_{e\gamma}}\,\,\Big[ {3 \over 4}(1-\mu^2)\Pi(K,\tau)\Big],
\end{equation}
 where
\begin{equation}
\Pi=\Delta_{T2}^{(S)}+\Delta_{P2}^{(S)}+\Delta_{P0}^{(S)}
\end{equation}
  The differential optical depth $\dot\tau_{e\gamma}(\tau)$ and total optical depth $\tau_{e\gamma}(\tau)$ due to the Thomson scattering at time  $\tau$ are defined as
\begin{equation}\label{optical}
    \dot{\tau}_{e\gamma}=a\,n_e\,\sigma_T,\,\,\,\,\,\,\,\tau_{e\gamma}(\tau)=\int_\tau^{\tau_0}\dot{\tau}_{e\gamma}(\tau) d\tau.
\end{equation}
As is shown in (\ref{Boltzmann}), the temperature  anisotropy
$\Delta_I^{(S)}$ doesn't have any source due to the forward
Compton scattering  in the NC space-time therefore, we only focus on the other equations to explore the NC effects.  Meanwhile, (\ref{Boltzmann03}) and
(\ref{Boltzmann3}) indicate that the effect of non-commutativity on the linear and circular polarization can be valuable for a significant value of $\frac{\kappa_{NC}}{\dot\tau_{e\gamma}}$ which is defined as follows
\begin{equation}\label{eq1}
  \tilde{\kappa}=  \frac{\kappa_{NC}}{\dot\tau_{e\gamma}}=\frac{3}{4}\frac{1}{\alpha}\,\frac{m_e^2}{\Lambda^2}\sum_{f=e,p}\frac{m_f}{k^0},
\end{equation}
which leads to larger values for Protons than the electrons.
\section{CMB power spectrum in NC space-time}
In the preceding section we found that the Compton scattering in the NC space changes the Boltzmann equations for the time evolution of the polarized CMB photons.  Here, we are ready to find the power spectra of $I$, $B$, $E$ and $V$ in the NC background.  To this end, we consider the power spectrum as
\begin{equation}\label{PS1}
    C_{Xl}=\frac{1}{2l+1}\sum_m\Big<a^*_{X,lm}\,a_{X,lm}\Big>,\,\,\,\,\,\, X=\{I,E,B,V\},
\end{equation}
where
\begin{eqnarray}
  a_{E,lm} &=& -(a_{2,lm}+a_{-2,lm})/2 ,\label{ae}\\
  a_{B,lm} &=& i(a_{2,lm}-a_{-2,lm})/2 ,\label{ab} \\
  a_{V,lm} &=& \int\,d\Omega Y^*_{lm} \Delta_V, \label{av}
\end{eqnarray}
which for the circularly polarized part of the CMB photons by using (\ref{Boltzmann3}) in the power spectrum $C_{Vl}$, one has
 \begin{eqnarray}
 & &  C_{Vl}=\frac{1}{2l+1}\sum_m\Big<a^*_{V,lm}\,a_{V,lm}\Big>, \label{CVl}\\
  &\approx&\frac{1}{2l+1}\int d^3KP_{v}(\bf{K})\sum_m\Big|\int d\Omega Y^*_{lm}\int_0^{\tau_0} d\tau\,
\dot\tau_{e\gamma}\,e^{ix \mu -\tau_{e\gamma}}\,A\,\,\tilde{\kappa}\,\Delta _{P}^{(S)}\Big|^2 ,
\end{eqnarray}
where
\bea
& & P_{v}(\bf{K})\delta(\bf{K'}-\bf{K})=\Big<(\xi(\bf{K})v_b)(\xi(\bf{K'})v_b)\Big>,
\eea
and $P_{v}(K)$ is the velocity power spectrum  which can
be expressed in terms of the primordial scalar spectrum $P_{\phi}^{(S)}$ as \cite{book-gor}
\begin{equation}\label{PSV}
  P_{v}(\bf{K},\tau)\sim P_{\phi}^{(S)}(\bf{K},\tau).
\end{equation}
Now (\ref{CVl}) and (\ref{PSV}) can provide an estimate on  $C_{Vl}$ in terms of the linearly polarized power spectrum $C_{Pl}$ as follows
\begin{equation}\label{CVl1}
   \tilde{\kappa}_{min}^2\, C_{Pl}\,\le C_{Vl}\, \le \,\tilde{\kappa}_{max}^2\, C_{Pl},
\end{equation}
where
\begin{eqnarray}
   \tilde\kappa_{max}&=&\frac{3}{4}\frac{m_e+m_p}{T^0}\frac{1}{\alpha}\,\frac{m_e^2}{\Lambda^2}\simeq\, 1 (10TeV/\Lambda)^2, \\
 \tilde\kappa_{min}&=&\frac{3}{4}\frac{m_e+m_p}{T^{lss}}\frac{1}{\alpha}\,\frac{m_e^2}{\Lambda^2}\simeq\, 10^{-3}\, (10TeV/\Lambda)^2,\label{kappa-min}
\end{eqnarray}
in which $k^0=T^0$  and $k^0=T^{lss}$ are the energies of the CMB photons  at the present time and the last scattering epoch, respectively. Using  the experimental value for the linearly polarized power spectrum of the CMB photons which is of the order of $0.1\mu K^2$ for $l<250$ \cite{plank}, one finds from (\ref{CVl1})-(\ref{kappa-min}) an estimation on the range of $C_{Vl}$ as
\begin{equation}\label{CVl2}
  0.1 nK^2\,\le C_{Vl}\, \le \,\,0.1\mu K^2,
\end{equation}
for a conservative value of the NC scale of the order of
$\Lambda\sim10TeV$.  Meanwhile, for the more or less accepted value  $\Lambda\sim 1TeV$  \cite{Haghighat}, the circular polarization power spectrum $C_{Vl}$  can be obtained in a range as follows
\begin{equation}\label{CVl2}
   10^{-3}\mu K^2\,\le C_{Vl}\, \le \,\,10^3\mu K^2,
\end{equation}
which is in the range of achievable experimental values.

In addition to  $C_{Vl}$ the Compton scattering in the NC space, in contrast with the ordinary space, can also generate the B-mode polarization.
To explore such a property we give the CMB
polarization  in terms of the divergence-free part (B-mode
$\Delta_{B}^{(S)}$) and the curl-free part (E-mode $\Delta_{E}^{(S)}$)
which are defined as
\begin{eqnarray}\label{Emode}
\Delta_{E}^{(S)}(\hat{n})&\equiv&-\frac{1}{2}[\bar{\eth}^{2}\Delta_{P}^{+(S)}(\hat{\bf{n}})+\eth^{2}\Delta_{P}^{-(S)}(\hat{\bf{n}})],\\
\label{Bmode}\Delta_{B}^{(S)}(\hat{n})&\equiv&\frac{i}{2}[\bar{\eth}^{2}\Delta_{P}^{+(S)}(\hat{\bf{n}})-\eth^{2}\Delta_{P}^{-(S)}(\hat{\bf{n}})],
\end{eqnarray}
where $\eth$ and $\bar{\eth}$ are spin raising and lowering
operators respectively \cite{hu}. As Eqs.(\ref{Boltzmann3}), (\ref{Bmode}) and (\ref{PS1}) show  the B-mode
 power spectrum $C_{Bl}$ due to the forward
electron and proton Compton scattering in the NC space-time depends on
the circular polarization power spectrum which can be estimated as
\begin{equation}\label{Bmode1}
    C^S_{Bl}\propto \bar{\tilde\kappa}^2 C_{Vl},\,\,\,\,\,\,\tilde\kappa_{min}\, <\,\bar{\tilde\kappa}\,<\,\tilde\kappa_{max},
\end{equation}
where $S$ indicates the scalar mode of the matter perturbation. Furthermore, the B-mode power spectrum depends on the scale of NC parameter $\Lambda$, through $\bar{\tilde\kappa} $ which can have a significant effect
on the value of the $r$-parameter even for $\Lambda\sim 10 TeV$.
In fact, by using Eqs. (\ref{CVl1}) and (\ref{Bmode1}) one has
\begin{equation}\label{Bmode2}
   \tilde{\kappa}_{min}^4\, C_{Pl}\,\le C^S_{Bl}\, \le \,\tilde{\kappa}_{max}^4\, C_{Pl},
\end{equation}
where for $\Lambda\sim 10TeV$ leads to
\begin{equation}\label{Bmode3}
   0.1 pK^2\,\le C^S_{Bl}\, \le \,0.1\mu K^2,
\end{equation}
and
\begin{equation}\label{Bmode4}
  10~nK^2\,\le C^S_{Bl}\, \le \,10~m K^2,
\end{equation}
for  $\Lambda\sim 1TeV$ which is comparable with the value of $C^{ob}_{Bl}\sim 0.01\mu
K^2$  for $l<250$ \cite{plank}.
\section{Conclusion}
We considered the time evolution of the Stokes parameters in the
NC-space-time. We showed that the NC corrections on the Compton
scattering can lead to the circular polarization for the CMB
radiation. It is also shown that the B-mode spectrum in contrast
with the usual production via the tensor mode perturbation,
 can be generated by the scalar mode perturbation in the
 NC-space-time.  The obtained result shows that
to fully understand the origin of the reported r parameter
\cite{BICEP2,plank}, one should consider all the alternative
sources for the B-mode spectrum of polarized CMB photons.
Furthermore, we found the V-mode power spectrum in the range of
Nano-kelvin squared and higher for the NC-scale $1TeV$ to $10TeV$
which is in the range of the accuracy of observational data
\cite{plank,act,pix,spider}.  We also showed that  the B-mode
 power spectrum $C_{Bl}$ due to the forward
electron and proton Compton scattering in the NC space-time depends on
the circular polarization power spectrum for the scalar mode of the matter perturbation.



\begin{thebibliography}{99}
\bibitem{jackson}
J.~D.~Jackson,{\it "Classical Electrodynamic"}, Wiley and Sons: New York(1998).

\bibitem{kosowsky}
 A.~Kosowsky,
  Annals Phys.\  {\bf 246}, 49 (1996)
  [arXiv:9501045[astro-ph]].

\bibitem{Hu}
  W.~Hu and M.~J.~White,
  New Astron.\  {\bf 2}, 323 (1997),
  [arXiv:9706147[astro-ph]].

\bibitem{Cooray:2002nm}
  A.~Cooray, A.~Melchiorri and J.~Silk,
  Phys.\ Lett.\  B {\bf 554}, 1 (2003) [arXiv:0205214[astro-ph]].

\bibitem{gio1}
M.~Giovannini,  (2002), [arXiv:0208152[hep-ph]].

\bibitem{gio2}
M.~Giovannini and K.~E.~Kunze, Phys.\ Rev.\ D {\bf 78}, 023010 (2008),
[arXiv:0804.3380[astro-ph]].

\bibitem{cmbpol}
E.~bavarsad, M.~Haghighat, Z.~Rezaei, R.~Mohammadi, I.~Motie, M.~Zarei
 Phys.\ Rev.\  D {\bf 81}: 084035 (2010)
  [arXiv:0912.2993[hep-th]].

\bibitem{Roh}
 R. Mohammadi, Eur.\ Phys.\ J.\ C {\bf 74}, no. 10, 3102 (2014) [arXiv:1312.2199[astro-ph.CO]];  R.~Mohammadi and S.-S.~Xue,
  Phys.\ Lett.\ B {\bf 731}, 272 (2014)
  [arXiv:1312.3862[hep-ph]].

\bibitem{xue}
I.~Motie and S.~-S.~Xue,
  Europhys.\ Lett.\  {\bf 100}, 17006 (2012)
  [arXiv:1104.3555[hep-ph]];
  R.~F.~Sawyer,
  [arXiv:1205.4969[astro-ph.CO]].

   \bibitem{BICEP2}
  P.~A.~R.~Ade {\it et al.}  [BICEP2 Collaboration],
 [arXiv:1403.3985[astro-ph.CO]].
\bibitem{keck}
  P.~A.~R.~Ade {\it et al.} [BICEP2 and Keck Array Collaborations],
  arXiv:1510.09217 [astro-ph.CO].
 \bibitem{14}
C. Scoccola, D. Harari and S. Mollerach,  Phys.\ Rev.\ D {\bf 70}, 063003 (2004); L. Cam-
panelli, A. D. Dolgov, M. Giannotti and F. L. Villante, Astrophys.\ J.\ {\bf 616}, 1 (2004);
A. Kosowsky, T. Kahniashvili, G. Lavrelashvili and B. Ratra, Phys.\ Rev.\ D {\bf 71}, 043006
(2005); M. Giovannini, Phys.\ Rev.\ D {\bf 71}, 021301 (2005); M. Giovannini and K. E. Kunze,
 Phys.\ Rev.\ D {\bf 78}, 023010 (2008);  Phys.\ Rev.\ D {\bf 79}, 063007 (2009); L. Pogosian,
A. P. S. Yadav, Y. -F. Ng and T. Vachaspati,  Phys.\ Rev.\ D {\bf 84}, 043530 (2011); M. Gio-
vannini,  Phys.\ Rev.\ D {\bf 89}, 061301 (2014); C. Bonvin, R. Durrer and R. Maartens,Phys.\ Rev.\ Lett.{\bf 112}, 191303 (2014)
[arXiv:1403.6768[astro-ph.CO]].

\bibitem{Giovannini}
  M.~Giovannini,
   Phys.\ Rev.\ D {\bf 90},  041301 (2014), [arXiv:1404.3974 [astro-ph.CO]].

  \bibitem{roh}  J.~Khodagholizadeh, R.~Mohammadi and S-S.~Xue,
  Phys.\ Rev.\ D {\bf 90}, 091301 (2014),[arXiv:1406.6213[astro-ph.CO]]; R.~Mohammadi, J.~Khodagholizadeh, M.~Sadegh and S.~S.~Xue,
  [arXiv:1602.00237 [astro-ph.CO]].

\bibitem{20Tev}
  P.~K.~Joby, P.~Chingangbam and S.~Das,
  Phys.\ Rev.\ D {\bf 91}, no. 8, 083503 (2015), [arXiv:1412.6036[astro-ph.CO]].

\bibitem{ardalan}
F. Ardalan, H. Arfaei and M. M. Sheikh-Jabbari, JHEP 9902 (1999) 016, [arXiv:9810072[hep-th]].

\bibitem{seiberg}
N. Seiberg and E. Witten, JHEP 9909 (1999) 032, [arXiv:9908142[hep-th]].

 \bibitem{Calmet}
 X.~Calmet, B.~Jurco, P.~Schupp, J.~Wess and M.~Wohlgenannt,
 Eur.\ Phys.\ J.\ C {\bf 23}, 363 (2002).

\bibitem{ncf2}
M. Chaichian, P. Presnajder, M. M. Sheikh-Jabbari and A. Tureanu,
[arXiv:0107037[hep-th]]; M. Chaichian, P. Presnajder,
M. M. Sheikh-Jabbari and A. Tureanu,[arXiv:0107055[ hep-th]]; I. F. Riad
and M. M. Sheikh-Jabbari, JHEP  0008, 045 (2000)[arXiv:0008132[hep-th]];  M.
Chaichian, M. M. Sheikh-Jabbari and A. Tureanu, Phys.\ Rev.\ Lett.\
{\bf 86}, 2716 (2001).

\bibitem{melic}
B. ~Melic, K. ~Passek-Kumericki, J. ~Trampetic, P. ~Schupp, M. ~Wohlgenannt
Eur.\ Phys.\ J.\ C {\bf 42} (2005) 483,
[arXiv:0502249[hep-ph]].

\bibitem{zal} M. Zaldarriaga and U. Seljak,
 Phys.\ Rev.\ D {\bf 55}, 1830 (1997), [arXiv:9609170[astro-ph]];  M.~Zaldarriaga, D.~N.~Spergel and U.~Seljak,
  Astrophys.\ J.\ {\bf 488}, 1 (1997)
  [arXiv:9702157[astro-ph]].

\bibitem{chandra}
 S. Chandrasekhar, {\it "Radiative Transfer"}, Dover, New York, 1960.

\bibitem{kaiser}
N. Kaiser, Mon. Not. R. Astron. Soc. {\bf 202}, 1169 (1983).

\bibitem{hu}
 W. Hu and M. J. White,
 New Astron. {\bf2}, 323 (1997),
[arXiv:9706147[astro-ph]].

\bibitem{book-gor}
Gorbunov, D.S. and Rubakov, {\it "V.A.- Introduction to the Theory of the Early Universe Cosmological Perturbations and Inflationary Theory"} -World Scientific Publishing Company Incorporated (2011).

\bibitem{plank}
Planck Collaboration I,
(2013),  [planck2013-p01, p08];
 P.~A.~R.~Ade {\it et al.}  [Planck Collaboration],
  [arXiv:1502.02114 [astro-ph.CO]].

  \bibitem{Haghighat}
  M. Haghighat and M. Khorsandi, 
  Eur. Phys.\ J.\ C {\bf 75}, 4 (2015). [arXiv:1410.0836[hep-ph]];
  M.~Haghighat, N.~Okada and A.~Stern,
  Phys.\ Rev.\ D {\bf 82}, 016007 (2010)
  [arXiv:1006.1009[hep-ph]]; A. Prakash, A. Mitra and P. K. Das, Phys.\ Rev.\ D {\bf 82}, 055020 (2010); M. M. Ettefaghi,  Phys.\ Rev.\ D {\bf 86}, 085038 (2012).

\bibitem{act}
M.~D.~Niemack,P.~A.~R.~Ade,J.~Aguirre, et al., Proc. SPIE, Vol. 7741 (2010), [arXiv:1006.5049[astro-ph.IM]].

\bibitem{pix}
A. Kogut and et al., (2011), [arXiv:1105.2044[astro-ph.CO]].

\bibitem{spider}
J.P. Filippini , P.A.R. Ade, , et al., (2011), [arXiv:1106.2158[astro-ph.CO]].

\end{thebibliography}
\end{document}